\documentclass[nofootinbib,amsmath,amssymb,aps,floatfix,notitlepage,superscriptaddress,longbibliography]{revtex4-1}
\pdfoutput=1

\usepackage{geometry}                %
\usepackage[T1]{fontenc}
\usepackage[utf8]{inputenc}
\usepackage{subcaption}
\usepackage{tikz}
\usetikzlibrary{positioning}
\usetikzlibrary{patterns}
\usetikzlibrary{decorations.markings}

\tikzset{
    fermion/.style={postaction={decorate},
        decoration={markings, mark=at position .7 with {\arrow{>}}}
    },
        meson/.style={
            dash pattern=on 1pt off .5 pt
        }
}

\usepackage{slashed}

\usepackage[pdfauthor={Michael Borinsky, Gerald V. Dunne},
            pdftitle={Non-Perturbative Completion of Hopf-Algebraic Dyson-Schwinger Equations},
            pdfsubject={},
            pdfkeywords={Renormalons, Resurgence, non-perturbative, Quantum Field Theory, Renormalization, Hopf algebra}]{hyperref}

\newcommand{\bigO}{\mathcal{O}}

\newcommand{\R}{\mathbb{R}}

\newcommand{\asy}{\mathcal{A}}
\newcommand{\wt}{\widetilde}

\definecolor{red}{rgb}{1,0,0} 

\definecolor{grey}{rgb}{.5,.5,.5} 

\definecolor{darkgreen}{rgb}{0, .7, 0}

\definecolor{purple}{rgb}{.7, 0, 1}

\begin{document}

\title{Non-Perturbative Completion of Hopf-Algebraic Dyson-Schwinger Equations}

\author{Michael Borinsky}
\affiliation{Nikhef Theory Group, Amsterdam 1098 XG, The Netherlands}
\author{Gerald V. Dunne}
\affiliation{Department of Physics, University of Connecticut, Storrs CT 06269-3046, USA}

\vspace*{-2\baselineskip}%
\hspace*{\fill} \mbox{\footnotesize{\textsc{Nikhef 2020-016}}}

\begin{abstract} 
For certain quantum field theories, the Kreimer-Connes Hopf-algebraic approach to renormalization reduces the Dyson-Schwinger equations to a system of non-linear ordinary differential equations for the expansion coefficients of the renormalized Green's function.  We apply resurgent asymptotic analysis to find the trans-series solutions which provide the non-perturbative completion of these formal Dyson-Schwinger expansions. We illustrate the general approach with the concrete example of four dimensional massless Yukawa theory, connecting with the exact functional solution found by Broadhurst and Kreimer. The trans-series solution is associated with the iterative form of the Dyson-Schwinger equations, and displays renormalon-like structure of integer-repeated Borel  singularities. Extraction of the Stokes constant is possible due to a property  we call `functional resurgence'.
\end{abstract}

\maketitle
\section{Introduction}
\label{sec:intro}

The Kreimer-Connes approach to renormalization in quantum field theory recasts the perturbative renormalization process in Hopf-algebraic terms, leading to new perspectives as well as new computational methods \cite{Kreimer:1997dp,Connes:1999zw,Connes:2000fe,yeats2017combinatorial,borinsky2018graphs}. A long-standing problem is to understand how non-perturbative effects fit naturally into this formalism. 
In this paper we present an approach to this problem based on \'Ecalle's resurgent trans-series and alien calculus \cite{ecalle1981fonctions,costin2008asymptotics,mitschi2016divergent,Aniceto:2018bis}.
We illustrate the general method by considering a local quantum field theory with a Green's function depending on a single running coupling, $\alpha$, and a single kinematical variable, $L=\ln q^2/\mu^2$, where $\mu$ is the renormalization scale. It has been shown by Broadhurst and Kreimer \cite{Broadhurst:1999ys,Broadhurst:2000dq}, and Kreimer and Yeats \cite{Kreimer:2006ua,Kreimer:2006gm}, that the recursive Hopf-algebraic structure of the Dyson-Schwinger equations, combined with the renormalization group equations describing the anomalous scaling under re-scaling of parameters, reduces the problem to a set of non-linear ordinary differential equations (ODEs). %
This was used in \cite{Broadhurst:1999ys,Broadhurst:2000dq} to extend Dyson-Schwinger solutions well beyond the simple ``rainbow'' and ``chain'' approximations, to a Hopf-based solution which sums over all possible nestings and chainings of the one loop self-energy. This is a highly non-trivial implementation of the BPHZ renormalization procedure, which is brought under algebraic and combinatorial control through the asymptotics of the growth of skeleton graphs. In the breakthrough papers 20 years ago \cite{Broadhurst:1999ys,Broadhurst:2000dq}, this was implemented explicitly for four dimensional massless Yukawa theory, enabling a solution to $30^{\rm th}$ perturbative order, a computation requiring $\sim 10^{20}$ BPHZ subtractions. 
This solution agrees with numerical integration and also Borel resummation techniques \cite{Broadhurst:1999ys,Broadhurst:2000dq,jentschura2001improved}. The relevant expansion coefficients of the anomalous dimension are related to the combinatorial problem of counting {\it connected chord diagrams} \cite{Stein1978,Flajolet2000}, and to a functional approach based on the properties of the ring of formal divergent series \cite{borinsky2018generating,borinsky2018graphs}. 
This enumerative graph interpretation
 is known to be a non-$D$-finite combinatorial problem \cite{klazar2003non}.
In this paper we study the associated ODEs using resurgent asymptotics and alien calculus, complementary approaches which yield non-perturbative  trans-series solutions, whose expansions display familiar features of resurgence such as large-order/low-order relations. The trans-series solution also provides insight into the origin of renormalon-like Borel plane behaviour arising from the iteration of Feynman diagram structures. For other analyses of resurgence properties of renormalization group and Dyson-Schwinger equations see \cite{Bellon:2008zz,Bellon:2016mje,Bellon:2018lwy,Bersini:2019axn}. A  novel perspective on this Hopf algebra based approach to QFT was recently uncovered by Kr\"uger \cite{Kruger:2019tas}. 
Renormalons have also been studied recently using ideas from resurgence, in a wide variety of theories: see for example \cite{Anber:2014sda,Dunne:2016nmc,Maiezza:2018pkk,Marino:2019wra,Marino:2019eym,Marino:2019fvu,Ishikawa:2019oga,Dondi:2020qfj}, and references therein.

\section{Broadhurst-Kreimer solution for the massless Yukawa theory}
\label{sec:hopf-yukawa}

As a concrete example to illustrate the general approach, we consider four dimensional massless Yukawa theory: 
\begin{eqnarray}
{\mathcal L}= \frac12 (\partial \phi)^2 + i \bar{\psi} \slashed{\partial} \,\psi - g\, \bar{\psi}\,\sigma \, \psi
\label{eq:ly}
\end{eqnarray}
As in \cite{Broadhurst:1999ys,Broadhurst:2000dq} we consider the renormalized fermion self-energy
\begin{align}
\Sigma(q) :=
    \begin{tikzpicture}[baseline={([yshift=-.6ex]current bounding box.center)}]  \coordinate (in); \coordinate[right=.25 of in] (v1); \coordinate[right=.25 of v1] (m); \coordinate[right=.25 of m] (v2); \coordinate[right=.25 of v2] (out);  \filldraw[preaction={fill,white},pattern=north east lines] (m) circle(.25);  \filldraw (v1) circle(1pt); \filldraw (v2) circle(1pt);  \draw[fermion] (in) -- (v1); \draw[fermion] (v2) -- (out);     \end{tikzpicture}
\end{align}
and take all propagator self-insertions into account. This approach can be depicted via the Dyson-Schwinger equation,
\begin{align}
\label{eq:pic_dse}
\begin{tikzpicture}[baseline={([yshift=-.6ex]current bounding box.center)}]  \coordinate (in); \coordinate[right=.25 of in] (v1); \coordinate[right=.25 of v1] (m); \coordinate[right=.25 of m] (v2); \coordinate[right=.25 of v2] (out);  \filldraw[preaction={fill,white},pattern=north east lines] (m) circle(.25);  \filldraw (v1) circle(1pt); \filldraw (v2) circle(1pt);  \draw[fermion] (in) -- (v1); \draw[fermion] (v2) -- (out);     \end{tikzpicture}
=
\begin{tikzpicture}[baseline={([yshift=-.6ex]current bounding box.center)}]  \coordinate (in); \coordinate[right=.25 of in] (v1); \coordinate[right=.5 of v1] (v2); \coordinate[right=.25 of v2] (out);  \draw[white] (v1) arc (-180:0:.25);  \filldraw (v1) circle(1pt); \filldraw (v2) circle(1pt);  \draw[fermion] (in) -- (v1); \draw[fermion] (v1) -- (v2); \draw[fermion] (v2) -- (out);  \draw[meson] (v1) arc (180:0:.25);  \end{tikzpicture}
+
\begin{tikzpicture}[baseline={([yshift=-.6ex]current bounding box.center)}]  \coordinate (in); \coordinate[right=.25 of in] (v1); \coordinate[right=.25 of v1] (v2); \coordinate[right=.25 of v2] (m); \coordinate[right=.25 of m] (v3); \coordinate[right=.25 of v3] (v4); \coordinate[right=.25 of v4] (out);  \draw[white] (v1) arc (-180:0:.5);  \filldraw[preaction={fill,white},pattern=north east lines] (m) circle(.25);  \filldraw (v1) circle(1pt); \filldraw (v2) circle(1pt); \filldraw (v3) circle(1pt); \filldraw (v4) circle(1pt);  \draw[fermion] (in) -- (v1); \draw[fermion] (v1) -- (v2); \draw[fermion] (v3) -- (v4); \draw[fermion] (v4) -- (out);  \draw[meson] (v1) arc (180:0:.5);  \end{tikzpicture}%
+
\begin{tikzpicture}[baseline={([yshift=-.6ex]current bounding box.center)}]  \coordinate (in); \coordinate[right=.25 of in] (v1); \coordinate[right=.25 of v1] (v2); \coordinate[right=.25 of v2] (m1); \coordinate[right=.25 of m1] (v3); \coordinate[right=.25 of v3] (v4); \coordinate[right=.25 of v4] (m2); \coordinate[right=.25 of m2] (v5); \coordinate[right=.25 of v5] (v6); \coordinate[right=.25 of v6] (out);  \draw[white] (v1) arc (-180:0:.5); \draw[white] (v4) arc (-180:0:.5);  \filldraw[preaction={fill,white},pattern=north east lines] (m1) circle(.25); \filldraw[preaction={fill,white},pattern=north east lines] (m2) circle(.25); \draw[white] (v1) arc (-180:0:.875);  \filldraw (v1) circle(1pt); \filldraw (v2) circle(1pt); \filldraw (v3) circle(1pt); \filldraw (v4) circle(1pt); \filldraw (v5) circle(1pt); \filldraw (v6) circle(1pt);  \draw[fermion] (in) -- (v1); \draw[fermion] (v1) -- (v2); \draw[fermion] (v3) -- (v4); \draw[fermion] (v5) -- (v6); \draw[fermion] (v6) -- (out);  \draw[meson] (v1) arc (180:0:.875);  \end{tikzpicture}%
+
\cdots
-
\text{subtractions}
\end{align}
with the appropriate BPHZ subtractions indicated. 
Another way to describe the relevant set of graphs is to start with the one-loop graph $\begin{tikzpicture}[baseline={([yshift=-.6ex]current bounding box.center)}]  \coordinate (in); \coordinate[right=.25 of in] (v1); \coordinate[right=.5 of v1] (v2); \coordinate[right=.25 of v2] (out);  \draw[white] (v1) arc (-180:0:.25);  \filldraw (v1) circle(1pt); \filldraw (v2) circle(1pt);  \draw[fermion] (in) -- (v1); \draw[fermion] (v1) -- (v2); \draw[fermion] (v2) -- (out);  \draw[meson] (v1) arc (180:0:.25);  \end{tikzpicture}$ and add all possible iterated and multiple insertions of this graph into itself. The pictorial equation \eqref{eq:pic_dse} corresponds to the integral equation
\begin{align}
\Sigma(q) = \frac{\alpha}{\pi^2} \int d^4 k\frac{1}{ (q+k)^2} \left( \frac{1}{\slashed{k}} 
+ \frac{1}{\slashed{k}} \Sigma(k) \frac{1}{\slashed{k}}
+ \frac{1}{\slashed{k}} \Sigma(k) \frac{1}{\slashed{k}} \Sigma(k) \frac{1}{\slashed{k}}
+ \cdots \right)
-
\text{subtractions}
\end{align}
where $\alpha=(g/(4\pi))^2$. 
By Poincar\'e symmetry, the fermion self-energy $\Sigma(q)$ must be proportional to $\slashed{q}$. We define the scalar-valued function $\wt{\Sigma}(q^2)$, such that
$\Sigma(q) = \slashed{q}\, \wt{\Sigma}(q^2)$,
and the integral equation reduces to
\begin{align}
\wt{\Sigma}(q^2) = \frac{\alpha}{\pi^2} \int d^4 k\frac{q \cdot k}{k^2 (q+k)^2 (1 - \wt{\Sigma}(k^2))} 
-
\text{subtractions},
\end{align}
where the subtractions can be chosen such that the momentum subtraction renormalization condition $\wt{\Sigma}(\mu^2) =0$ is fulfilled.

Broadhurst and Kreimer \cite{Broadhurst:2000dq} solved this integral equation using Hopf-algebraic methods. For the anomalous dimension in the momentum subtraction scheme
\begin{align}
\wt\gamma(\alpha)=\frac{d}{d\ln q^2}\ln \left(1-\wt\Sigma(q^2)\right)\Bigg|_{q^2=\mu^2},
\label{eq:gamma}
\end{align}
they obtained the non-linear ODE,
\begin{align}
2 \wt\gamma &= - \alpha -\wt \gamma^2 + 2\alpha  \wt\gamma \frac{d}{d \alpha} \wt\gamma.
\label{eq:gamma_ode}
\end{align}

Subsequently, Kreimer and Yeats \cite{Kreimer:2006ua} confirmed and generalized this analysis, and uncovered a close relationship with \emph{connected chord diagrams}, as reviewed below in Section \ref{sec:asym}. Among other things, they established that the full Green's function can be  recovered from eq.~\eqref{eq:gamma_ode},
\begin{eqnarray}
\wt\Sigma(q^2)&=&-\sum_{j=1}^\infty \wt\gamma_j(\alpha)\, L^j
\label{eq:gl}
\\
&=&-\sum_{j=1}^\infty c_j(L)\, \alpha^j.
\label{eq:ga}
\end{eqnarray}
where we recall the notation: $L\equiv \ln \frac{q^2}{\mu^2}$.
The first term, $\wt\gamma_1(\alpha)$, is the anomalous dimension $\wt\gamma(\alpha)$. All higher coefficients are expressed recursively in term of $\wt\gamma_1(\alpha)$:
\begin{eqnarray}
\wt\gamma_k(\alpha)=\frac{1}{k} \wt\gamma_{1}(\alpha)\left(1-2 \alpha\, \partial_\alpha\right) \wt\gamma_{k-1}(\alpha), \quad k\geq 2.
\label{eq:gk}
\end{eqnarray}
This equation is an avatar of the renormalization group equation in the Hopf-algebra approach \cite{Kreimer:2006ua}. The solution to the self-inserted Yukawa self-energy problem is the simplest non-trivial example of a more general framework. See \cite{Yeats:2008zy,vanBaalen:2008tc,vanBaalen:2009hu,Bellon:2008zz,Bellon:2016mje,Marie:2012cc,Courtiel:2019dnq} for generalizations and \cite[Chap. 9]{yeats2017combinatorial} for a recent review. 
\section{Asymptotic formal perturbative series}
\label{sec:asym}

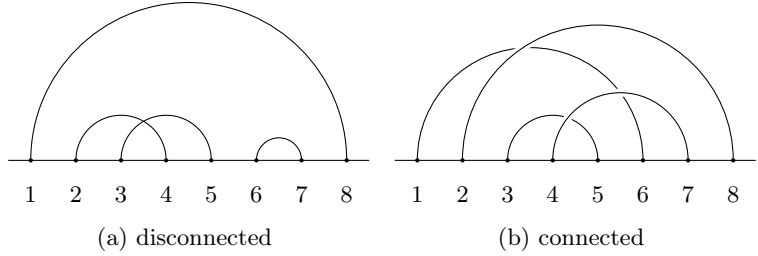
\begin{figure}
\begin{centering}
\subcaptionbox{disconnected\label{fig:disc_chord}}{
\begin{tikzpicture}[scale=0.6]    \draw (0,0) arc (0:-180:-3.500000);  \draw (1,0) arc (0:-180:-1.000000);  \draw (2,0) arc (0:-180:-1.000000);  \draw (5,0) arc (0:-180:-0.500000); \node at (0, -.75){$1$}; \draw[fill] (0, 0) circle (1pt); \node at (1, -.75){$2$}; \draw[fill] (1, 0) circle (1pt); \node at (2, -.75){$3$}; \draw[fill] (2, 0) circle (1pt); \node at (3, -.75){$4$}; \draw[fill] (3, 0) circle (1pt); \node at (4, -.75){$5$}; \draw[fill] (4, 0) circle (1pt); \node at (5, -.75){$6$}; \draw[fill] (5, 0) circle (1pt); \node at (6, -.75){$7$}; \draw[fill] (6, 0) circle (1pt); \node at (7, -.75){$8$}; \draw[fill] (7, 0) circle (1pt); \draw (-0.500000,0)--(7.500000,0); \end{tikzpicture}
}
\subcaptionbox{connected\label{fig:cntd_chord}}{
\begin{tikzpicture}[scale=0.6] \draw[color=white,line width=3pt] (0,0) arc (0:-180:-2.500000); \draw (0,0) arc (0:-180:-2.500000); \draw[color=white,line width=3pt] (2,0) arc (0:-180:-1.000000); \draw (2,0) arc (0:-180:-1.000000); \draw[color=white,line width=3pt] (1,0) arc (0:-180:-3.000000); \draw (1,0) arc (0:-180:-3.000000); \draw[color=white,line width=3pt] (3,0) arc (0:-180:-1.500000); \draw (3,0) arc (0:-180:-1.500000); \node at (0, -.75){$1$}; \draw[fill] (0, 0) circle (1pt); \node at (1, -.75){$2$}; \draw[fill] (1, 0) circle (1pt); \node at (2, -.75){$3$}; \draw[fill] (2, 0) circle (1pt); \node at (3, -.75){$4$}; \draw[fill] (3, 0) circle (1pt); \node at (4, -.75){$5$}; \draw[fill] (4, 0) circle (1pt); \node at (5, -.75){$6$}; \draw[fill] (5, 0) circle (1pt); \node at (6, -.75){$7$}; \draw[fill] (6, 0) circle (1pt); \node at (7, -.75){$8$}; \draw[fill] (7, 0) circle (1pt); \draw (-0.500000,0)--(7.500000,0); \end{tikzpicture}
}
\end{centering}
\caption{A disconnected and a connected chord diagram}
\label{fig:disc_ctnd_chord}
\end{figure}
Solving the non-linear ODE \eqref{eq:gamma_ode} iteratively in orders of $\alpha$ gives the formal perturbative expansion of the anomalous dimension:
\begin{eqnarray}
\wt\gamma(\alpha)\sim \sum_{n=1}^\infty C_n \frac{(-\alpha)^n}{2^{2n-1}}
= -\frac{\alpha}{2}+\frac{\alpha^2}{2^3}-4\frac{\alpha^3}{2^5}+27\frac{\alpha^4}{2^7}+\dots \text{ as } \alpha \rightarrow 0
\label{eq:hopf}
\end{eqnarray}
where the coefficients $C_n$ are generated by the recursion formula,
\begin{eqnarray}
C_{n+1}=n\sum_{k=1}^n C_n C_{n+1-k}, \quad n\geq 1\quad (C_1=1),
\label{eq:hcn}
\end{eqnarray}
This recursion relation enumerates \emph{connected chord diagrams} \cite{Kreimer:2006ua,Stein1978}. A chord diagram of order $n$ is a matching of $2n$ points. There are $(2n-1)!!$ chord diagrams. A chord diagram is \emph{connected} if there is no way to draw the diagram without crossing chords. See Fig.~\ref{fig:disc_ctnd_chord} for illustrations of a disconnected and a connected chord diagram.
The first terms are:
\begin{eqnarray}
C_n=\left[1, 1, 4, 27, 248, 2830, 38232, 593859, 10401712, 202601898, \dots\right]
\label{eq:cns}
\end{eqnarray}
This sequence is listed in the OEIS \cite{oeis} as \texttt{A000699}.
The numbers $C_n$ diverge factorially with order $n$, so the expansion (\ref{eq:hopf}) is a formal asymptotic series. The large-order behaviour of the $C_n$ is:
\begin{align}
\begin{aligned}
C_n\sim e^{-1} \frac{2^{n +\frac12 }\, \Gamma\left(n+\frac{1}{2}\right)}{\sqrt{2\pi}}\Big(
&
1-\frac{\frac{5}{2}}{ 2 \left(n-\frac{1}{2}\right)} -\frac{\frac{43}{8} }{2^2 \left( n-\frac{1}{2}\right)  \left(n-\frac{3}{2}\right)}
\\
&
-\frac{\frac{579}{16}}{ 2^3 \left(n-\frac{1}{2}\right) \left(n-\frac{3}{2}\right) \left(n-\frac{5}{2}\right)}-\dots \Big).
\label{eq:large-order}
\end{aligned}
\end{align}
The first coefficient of this asymptotic expansion was first evaluated by Kleitman \cite{kleitman1970proportions} and later confirmed by Stein and Everett \cite{Stein1978class}. The higher order corrections have been given in \cite{borinsky2018generating}.

\begin{figure}
\begin{centering}
\subcaptionbox{Rainbow approximation\label{fig:rainbow}}{
$
\begin{tikzpicture}[baseline={([yshift=-.6ex]current bounding box.center)}]  \coordinate (in); \coordinate[right=.25 of in] (v1); \coordinate[right=.25 of v1] (m); \coordinate[right=.25 of m] (v2); \coordinate[right=.25 of v2] (out);  \filldraw[preaction={fill,white},pattern=north east lines] (m) circle(.25);  \filldraw (v1) circle(1pt); \filldraw (v2) circle(1pt);  \draw[fermion] (in) -- (v1); \draw[fermion] (v2) -- (out);     \end{tikzpicture}=
\begin{tikzpicture}[baseline={([yshift=-.6ex]current bounding box.center)}]  \coordinate (in); \coordinate[right=.25 of in] (v1); \coordinate[right=.5 of v1] (v2); \coordinate[right=.25 of v2] (out);  \draw[white] (v1) arc (-180:0:.25);  \filldraw (v1) circle(1pt); \filldraw (v2) circle(1pt);  \draw[fermion] (in) -- (v1); \draw[fermion] (v1) -- (v2); \draw[fermion] (v2) -- (out);  \draw[meson] (v1) arc (180:0:.25);  \end{tikzpicture}
+
\begin{tikzpicture}[baseline={([yshift=-.6ex]current bounding box.center)}]  \coordinate (in); \coordinate[right=.25 of in] (v1); \coordinate[right=.25 of v1] (v2); \coordinate[right=.25 of v2] (m); \coordinate[right=.25 of m] (v3); \coordinate[right=.25 of v3] (v4); \coordinate[right=.25 of v4] (out);  \draw[white] (v1) arc (-180:0:.5);  \filldraw (v1) circle(1pt); \filldraw (v2) circle(1pt); \filldraw (v3) circle(1pt); \filldraw (v4) circle(1pt);  \draw[fermion] (in) -- (v1); \draw[fermion] (v1) -- (v2); \draw[fermion] (v2) -- (v3); \draw[fermion] (v3) -- (v4); \draw[fermion] (v4) -- (out);  \draw[meson] (v1) arc (180:0:.5); \draw[meson] (v2) arc (180:0:.25);  \end{tikzpicture}
+
\begin{tikzpicture}[baseline={([yshift=-.6ex]current bounding box.center)}]  \coordinate (in); \coordinate[right=.25 of in] (v1); \coordinate[right=.25 of v1] (v2); \coordinate[right=.25 of v2] (v3); \coordinate[right=.25 of v3] (m); \coordinate[right=.25 of m] (v4); \coordinate[right=.25 of v4] (v5); \coordinate[right=.25 of v5] (v6); \coordinate[right=.25 of v6] (out);  \draw[white] (v1) arc (-180:0:.75);  \filldraw (v1) circle(1pt); \filldraw (v2) circle(1pt); \filldraw (v3) circle(1pt); \filldraw (v4) circle(1pt); \filldraw (v5) circle(1pt); \filldraw (v6) circle(1pt);  \draw[fermion] (in) -- (v1); \draw[fermion] (v1) -- (v2); \draw[fermion] (v2) -- (v3); \draw[fermion] (v3) -- (v4); \draw[fermion] (v4) -- (v5); \draw[fermion] (v5) -- (v6); \draw[fermion] (v6) -- (out);  \draw[meson] (v1) arc (180:0:.75); \draw[meson] (v2) arc (180:0:.5); \draw[meson] (v3) arc (180:0:.25);  \end{tikzpicture}%
+
\cdots
$
\vspace{-3ex}
}
\subcaptionbox{Chain approximation\label{fig:chain}}{
$
\begin{tikzpicture}[baseline={([yshift=-.6ex]current bounding box.center)}]  \coordinate (in); \coordinate[right=.25 of in] (v1); \coordinate[right=.25 of v1] (m); \coordinate[right=.25 of m] (v2); \coordinate[right=.25 of v2] (out);  \filldraw[preaction={fill,white},pattern=north east lines] (m) circle(.25);  \filldraw (v1) circle(1pt); \filldraw (v2) circle(1pt);  \draw[fermion] (in) -- (v1); \draw[fermion] (v2) -- (out);     \end{tikzpicture}=
\begin{tikzpicture}[baseline={([yshift=-.6ex]current bounding box.center)}]  \coordinate (in); \coordinate[right=.25 of in] (v1); \coordinate[right=.5 of v1] (v2); \coordinate[right=.25 of v2] (out);  \draw[white] (v1) arc (-180:0:.25);  \filldraw (v1) circle(1pt); \filldraw (v2) circle(1pt);  \draw[fermion] (in) -- (v1); \draw[fermion] (v1) -- (v2); \draw[fermion] (v2) -- (out);  \draw[meson] (v1) arc (180:0:.25);  \end{tikzpicture}
+
\begin{tikzpicture}[baseline={([yshift=-.6ex]current bounding box.center)}]  \coordinate (in); \coordinate[right=.25 of in] (v1); \coordinate[right=.25 of v1] (v2); \coordinate[right=.25 of v2] (m); \coordinate[right=.25 of m] (v3); \coordinate[right=.25 of v3] (v4); \coordinate[right=.25 of v4] (out);  \draw[white] (v1) arc (-180:0:.5);  \filldraw (v1) circle(1pt); \filldraw (v2) circle(1pt); \filldraw (v3) circle(1pt); \filldraw (v4) circle(1pt);  \draw[fermion] (in) -- (v1); \draw[fermion] (v1) -- (v2); \draw[fermion] (v2) -- (v3); \draw[fermion] (v3) -- (v4); \draw[fermion] (v4) -- (out);  \draw[meson] (v1) arc (180:0:.5); \draw[meson] (v2) arc (180:0:.25);  \end{tikzpicture}
+
\begin{tikzpicture}[baseline={([yshift=-.6ex]current bounding box.center)}]  \coordinate (in); \coordinate[right=.25 of in] (v1); \coordinate[right=.25 of v1] (v2); \coordinate[right=.25 of v2] (m1); \coordinate[right=.25 of m1] (v3); \coordinate[right=.25 of v3] (v4); \coordinate[right=.25 of v4] (m2); \coordinate[right=.25 of m2] (v5); \coordinate[right=.25 of v5] (v6); \coordinate[right=.25 of v6] (out);  \draw[white] (v1) arc (-180:0:.875);  \filldraw (v1) circle(1pt); \filldraw (v2) circle(1pt); \filldraw (v3) circle(1pt); \filldraw (v4) circle(1pt); \filldraw (v5) circle(1pt); \filldraw (v6) circle(1pt);  \draw[fermion] (in) -- (v1); \draw[fermion] (v1) -- (v2); \draw[fermion] (v2) -- (v3); \draw[fermion] (v3) -- (v4); \draw[fermion] (v4) -- (v5); \draw[fermion] (v5) -- (v6); \draw[fermion] (v6) -- (out);  \draw[meson] (v1) arc (180:0:.875); \draw[meson] (v2) arc (180:0:.25); \draw[meson] (v4) arc (180:0:.25);  \end{tikzpicture}%
+
\cdots
$
\vspace{-3ex}
}
\end{centering}
\caption{Simpler approximations for the Yukawa fermion propagator} \label{fig:rainbow_chain_approximation}
\end{figure}
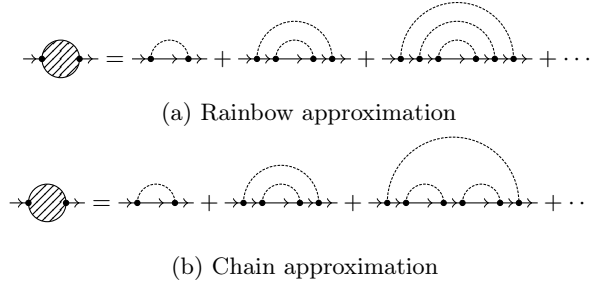

\begin{figure}[htb]
\begin{centering}
\includegraphics[scale=.7]{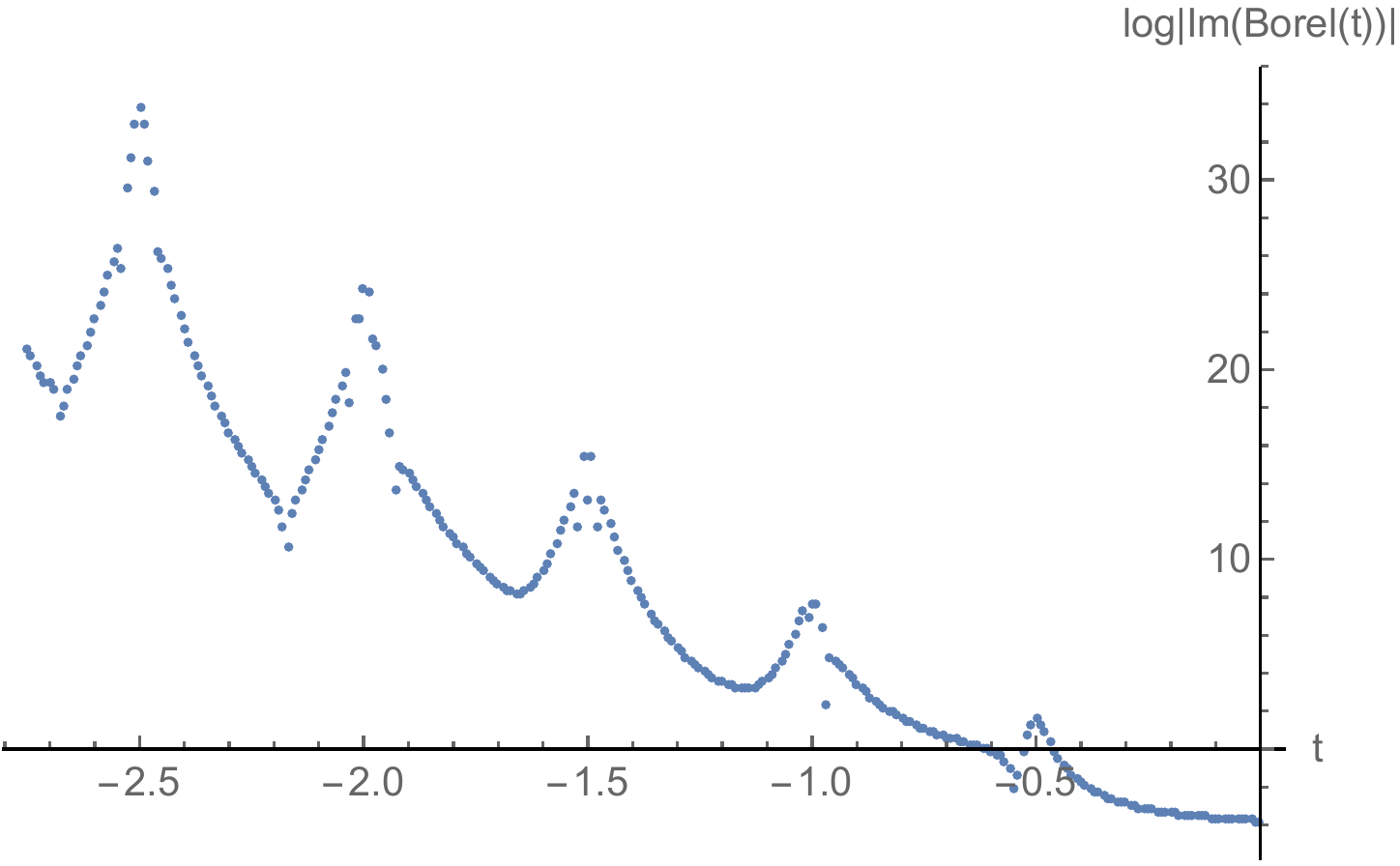}
\end{centering}
\caption{Plot of the Borel plane singularity structure for the Borel transform of the anomalous dimension, for the 4d massless Yukawa theory. The natural logarithm of the absolute value of the imaginary part of the Borel transform (constructed via a Pad\'e approximant in the conformally mapped disc  \cite{costin2008asymptotics}) is plotted for $t$ just above the negative real Borel axis: $t\to t+\frac{i}{50}$.
We clearly see a leading singularity on the negative Borel axis at $t=-\frac{1}{2}$, and further Borel singularities at integer-multiples of this leading location. These singularities correspond directly to the higher non-perturbative terms in the trans-series, and are associated with the iterative renormalon-like structure of the Dyson-Schwinger equations.}
\label{fig:borel}
\end{figure}

It is interesting to compare the Hopf-algebraic perturbative expansion (\ref{eq:hopf}) with two simpler approximations to the Dyson-Schwinger equations: the rainbow and the chain approximations, depicted in Fig.~\ref{fig:rainbow_chain_approximation}.
The rainbow approximation \cite{Delbourgo:1996vn} yields a {\it convergent} expansion of the  anomalous dimension
\begin{eqnarray}
\wt\gamma_{\rm rainbow}(\alpha)=1-\sqrt{1+\alpha}=-\frac{\alpha}{2}+\frac{\alpha^2}{2^3}-2\frac{\alpha^3}{2^5}+5\frac{\alpha^4}{2^7}+\dots
\label{eq:rainbow}
\end{eqnarray}
while the chain approximation \cite[eq.~(19)]{Broadhurst:1999ys} yields an {\it asymptotic} expansion
\begin{align}
\begin{aligned}
\wt\gamma_{\rm chain}(\alpha)=
-2\int_0^\infty \frac{dt}{t+1} \, e^{-4t/\alpha} 
&\sim 
-\frac{\alpha}{2}+\frac{\alpha^2}{2^3}-2\frac{\alpha^3}{2^5}+6\frac{\alpha^4}{2^7}+\dots\\
&\sim  \sum_{n=1}^\infty (-1)^n (n-1)! \frac{\alpha^n}{2^{2n-1}}.
\label{eq:chain}
\end{aligned}
\end{align}
The chain approximation is divergent, and has just one singularity in the Borel plane, a simple pole at $t=-1$, viewing the expansion (\ref{eq:chain}) as an expansion in $\alpha/4$. By contrast, the Hopf-algebraic result in (\ref{eq:hopf})-(\ref{eq:large-order}), which sums over all nestings and chainings of the one loop self-energy divergence, yields a divergent expansion whose Borel structure is much richer, revealing the characteristic renormalon-like structure of Borel singularities repeated at integer multiples of the leading singularity \cite{tHooft:1977xjm,Beneke:1998ui}. From the large-order growth (\ref{eq:large-order}), we see that the leading Borel singularity is at $t=-1/2$ (with the same normalization), and that it is a branch point rather than a pole. This situation is consistent with the Borel structure of zero dimensional Yukawa theory, where also a Borel singularity at $t=-\frac12$ is observed \cite[Sec.~6.4.3]{Borinsky:2017hkb}. In the four dimensional Yukawa theory treated here, further singularities appear on the negative Borel axis, at all integer multiples of the location of the leading singularity. See Fig.~\ref{fig:borel}. 
This integer-repetition of Borel singularities is a key indicator of non-perturbative physics, and can be identified using the Pad\'e-Conformal-Borel method as discussed in \cite{Costin:2020hwg}. %
Such a Borel plane singularity structure arises naturally in the context of non-linear differential equations \cite{costin2008asymptotics}, as we illustrate in Section \ref{sec:ode} of this paper. It also appears naturally in the framework of alien calculus, as shown in Section \ref{sec:alien}.
Ultimately it is associated with the iterative structure of the Dyson-Schwinger equations.
It is quite remarkable that the Hopf-algebraic solution of the four dimensional massless Yukawa theory is simple enough to be analyzable, and yet sophisticated enough to display non-trivial non-perturbative behaviour.

\section{Trans-series Solution from the Hopf Algebraic Dyson-Schwinger Differential Equation}
\label{sec:ode}

\subsection{Formal Perturbative Series}

The coefficients $C_n$ in (\ref{eq:hopf}) are generated through the generating function of connected chord diagrams  (see A000699 in the OEIS \cite{oeis}):
\begin{eqnarray}
C(x)=\sum_{n=1}^\infty C_n\, x^n
\label{eq:c-exp}
\end{eqnarray}
The correspondence with the anomalous dimension is:
\begin{eqnarray}
\wt\gamma(\alpha)=2\, C\left(-\frac{\alpha}{4}\right)
\label{eq:gamma-c}
\end{eqnarray}
The connected chord diagram generating function $C(x)$ satisfies the non-linear ordinary differential equation
\begin{eqnarray}
C(x)\left(1-2x\frac{d}{dx}\right)C(x)=x-C(x)
\label{eq:c}
\end{eqnarray}
which is a re-writing of eq.~\eqref{eq:gamma_ode}, or can be derived  
from the recurrence relation in eq.~\eqref{eq:hcn}. The differential equation (\ref{eq:c}) was also considered by Stein  \cite{Stein1978}, who established that its solution indeed enumerates connected chord diagrams. 

By making the ansatz $C(x) = c \, x^b + \bigO(x^{b+1})$ it is easy to confirm that eq.~\eqref{eq:c} has a unique solution $C \in \R[[x]]$ of the form $C(x) = x + \bigO(x^2)$. The reason for this is the irregular singular point of the differential equation at $x=0$. The free parameter, necessary for the solution of a first order differential equation, is hidden in \emph{non-perturbative corrections} to the power series solution. 

To go beyond the divergent perturbative expansion in (\ref{eq:c-exp}), we write a trans-series ansatz \cite{costin2008asymptotics}:
\begin{eqnarray}
C(x)=\sum_{k=0}^\infty \sigma^k C^{(k)}(x)
\label{eq:c-trans}
\end{eqnarray}
Here $C^{(0)}(x)$ is the formal perturbative series in (\ref{eq:c-exp}), and $C^{(k\geq 1)}(x)$ are exponentially small (as $x\to 0^+$) non-perturbative terms. The constant $\sigma$ in (\ref{eq:c-trans}) is  the  ``trans-series parameter'', or ``instanton counting parameter'', which encodes the boundary condition parameter for the ODE. Inserting this ansatz for $C(x)$ into  (\ref{eq:c}), and collecting powers of $\sigma$,  we obtain a tower of {\it linear} inhomogeneous equations for the non-perturbative terms $C^{(k\geq 1)}(x)$.

\subsection{One-Instanton Term}

The $\bigO(\sigma)$ term yields a {\it linear} inhomogeneous equation for the first non-perturbative term which can be expressed as:
\begin{eqnarray}
\frac{{C^{(1)}}^\prime(x)}{C^{(1)}(x)}=-\frac{{C^{(0)}}^\prime(x)}{C^{(0)}(x)} +\frac{1}{x}+\frac{1}{2 x\, C^{(0)}(x)}
\label{eq:c1np-eq}
\end{eqnarray}
from which we learn that
\begin{eqnarray}
C^{(1)}(x)=\frac{x}{C^{(0)}(x)}\exp\left[\int^x \frac{dt}{2 t\,  C^{(0)}(t)}\right]
\label{eq:c1np}
\end{eqnarray}
The overall integration constant can be absorbed into the trans-series parameter $\sigma$. We can simplify the expression (\ref{eq:c1np}) using the interesting identity
\begin{eqnarray}
\frac{1}{x\, C^{(0)}(x)}=-\frac{d}{dx}\left(\frac{\left(C^{(0)}(x)+1\right)^2}{x}+\ln x\right)
\label{eq:id1}
\end{eqnarray}
which follows from the non-linear ODE (\ref{eq:c})  satisfied by  $C^{(0)}(x)$. Thus we can write the one-instanton term $C^{(1)}(x)$ as:
\begin{eqnarray}
C^{(1)}(x)= \frac{1}{\sqrt{2\pi}} \frac{\sqrt{x}}{C^{(0)}(x)} \, \exp\left[-\frac{\left(C^{(0)}(x)+1\right)^2}{2x}\right]
\label{eq:c1}
\end{eqnarray}
where we have chosen the (arbitrary) overall normalization constant for later convenience: see (\ref{eq:asymptotic_C}) and (\ref{eq:iden}). Once we have made this normalization choice, it  propagates through all orders of the trans-series expansion (\ref{eq:c-trans}).

Using  the formal series for $C^{(0)}(x)$ we obtain the formal series expansion for $C^{(1)}(x)$:
\begin{eqnarray}
C^{(1)}(x)&\sim&\frac{e^{-1/(2x)}}{\sqrt{x}} \frac{e^{-1}}{\sqrt{2\pi}}   \left[1-\frac{5 x}{2}-\frac{43 x^2}{8}-\frac{579 x^3}{16}-\dots \right] 
\label{eq:c1np-exp}
\\
&\equiv & \frac{e^{-1/(2x)}}{\sqrt{x}} \sum_{n=0}^\infty C_n^{(1)} \, x^n
\label{eq:c1np-cs}
\end{eqnarray}
The first factor in (\ref{eq:c1np-cs}) is identified as the ``one-instanton'' factor
\begin{eqnarray}
\xi(x)\equiv \frac{e^{-1/(2x)}}{\sqrt{x}},
\label{eq:inst}
\end{eqnarray}
and the second factor is identified as the ``fluctuation about the one-instanton'', 
\begin{eqnarray}
C^{(1)}_{\rm fluc}(x)\equiv \sum_{n=0}^\infty C_n^{(1)} \, x^n.
\end{eqnarray}
This result for $C^{(1)}(x)$ exhibits two manifestations of resurgence. First, compare the coefficients $C_n^{(1)}$ in  (\ref{eq:c1np-exp}), of the fluctuation series $C^{(1)}_{\rm fluc}(x)$, with the coefficients in the large-order behaviour  of the coefficients of the formal perturbative series $C^{(0)}(x)$ in (\ref{eq:large-order}): $\left[1, -\frac{5}{2}, -\frac{43}{8}, -\frac{579}{16}, \dots\right]$. The coefficients of the large-order behaviour (\ref{eq:large-order}) of the perturbative series coefficients re-appear (``resurge'') as the coefficients of the fluctuations about the one-instanton term in  (\ref{eq:c1np-exp}). This is an example of the generic Berry-Howls large-order/low-order resurgence relation \cite{BerryHowls}.  But, more deeply, we see that the relation between the one-instanton term and the perturbative series is more explicit: expression (\ref{eq:c1}) shows that the one-instanton term $C^{(1)}(x)$ is explicitly encoded in terms of the ``zero-instanton'' term $C^{(0)}(x)$.

Note that in (\ref{eq:c1}) we have made a convenient choice for the arbitrary normalization of $C^{(1)}(x)$. Our choice is motivated by the overall multiplicative factor in the large-order growth (\ref{eq:large-order}), but we emphasize that while the normalization of $C^{(1)}(x)$ is arbitrary, the overall normalization of the large-order growth on the other hand is not arbitrary: it is a property of the differential equation, known as a \emph{Stokes constant}, and can be found by solving an associated \emph{connection problem}. See \cite{costin1998,costin2008asymptotics,Aniceto:2013fka,Aniceto:2018bis} for in depth accounts of Stokes constants and associated phenomena. It is in general a difficult problem to determine these Stokes constants, but for this specific problem it is possible to fix the normalization and determine the Stokes constant by using more sophisticated tools from \emph{alien calculus}. We  illustrate these tools and determine the normalization factor in Section~\ref{sec:alien}. In this example it is simple to generate very high orders of the perturbative expansion, so the Stokes constant can also be deduced {\it numerically} from the large order growth of the $C_n$ generated by (\ref{eq:hcn}).

 Similarly, we expect to find a relation between the one-instanton term $C^{(1)}(x)$ and the higher order instanton terms $C^{(k\geq 2)}(x)$ in the trans-series (\ref{eq:c-trans}). To this end, we can use the expression (\ref{eq:c1}) to   generate straightforwardly 
  the expansion coefficients of $C^{(1)}_{\rm fluc}(x)$ to very high order, from which we can determine the large-order growth of these coefficients, including sub-leading terms:
\begin{eqnarray}
C_n^{(1)}&\sim &
-2 e^{-2}\, \frac{2^{n +\frac32 }\, \Gamma\left(n+\frac{3}{2}\right)}{2\pi}
\left(1-\frac{5}{
   2\left(n+\frac{1}{2}\right)} - \frac{\frac{11}{2}}{2^2 \left(n+\frac{1}{2}\right) \left(n-\frac{1}{2}\right)}
    \right.\nonumber\\
   &&
   \left. 
   -\frac{\frac{97}{2}}{2^3 \left(n+\frac{1}{2}\right)\left(n-\frac{1}{2}\right) \left(n-\frac{3}{2}\right)}-\dots \right).
\label{eq:large-order-np1}
\end{eqnarray}

\subsection{Two-Instanton Term}

At order $\sigma^2$, the trans-series ansatz (\ref{eq:c-trans}) in the ODE (\ref{eq:c}) also yields a  linear inhomogeneous equation for the ``two-instanton'' term, $C^{(2)}(x)$, which can be written as
\begin{eqnarray}
2 x\, C^{(0)}\, {C^{(2)}}^\prime+(-1-2 C^{(0)}+2x \,{C^{(0)}}^\prime) C^{(2)}=C^{(1)}\left(C^{(1)}-2x\,  {C^{(1)}}^\prime\right)
\label{eq:c2np-eq}
\end{eqnarray}
Using the equations for $C^{(0)}(x)$ and $C^{(1)}(x)$, this simplifies to the following compact form:
\begin{eqnarray}
\left(\frac{C^{(2)}}{C^{(1)}}\right)^\prime = -\frac{1}{2}\frac{C^{(1)}}{(C^{(0)})^3}
\label{eq:cnp2}
\end{eqnarray}
Using further identities for $C^{(0)}(x)$ and $C^{(1)}(x)$ we can write the right-hand-side as a total derivative. To see this, define the following bivariate function:
\begin{eqnarray}
f(x,y)\equiv \frac{1}{\sqrt{2\pi}} \frac{x}{y} \exp\left[-\frac{1}{2x}\, y(y+2)\right] .
\label{eq:fxc}
\end{eqnarray}
In terms of this function $f(x,y)$, the one-instanton result (\ref{eq:c1}) can be written as
\begin{eqnarray}
C^{(1)}(x)= \xi(x) \cdot f(x, C^{(0)}(x))
\label{eq:c1f}
\end{eqnarray}
where we recall that $\xi(x)$ is the non-perturbative instanton factor defined in (\ref{eq:inst}).
Furthermore, the differential equations for $C^{(0)}(x)$ and $C^{(1)}(x)$ imply that 
\begin{eqnarray}
\frac{C^{(1)}}{(C^{(0)})^3}=- \frac{d}{dx}\left(\xi(x) \left[\frac{\partial f(x,y)}{\partial y}\right]_{y=C^{(0)}(x)}\right)
\label{eq:id2}
\end{eqnarray}
Consequently, we can write $C^{(2)}(x)$ as follows:
\begin{eqnarray}
C^{(2)}(x)= \frac{1}{2!} \xi(x)^2 \cdot f(x, C^{(0)}(x))\left[ \frac{\partial f(x, y)}{\partial y}\right]_{y=C^{(0)}(x)}
\label{eq:cnp22}
\end{eqnarray}
Note that expression (\ref{eq:c2np-eq}) is an inhomogeneous equation, so the normalization of $C^{(2)}(x)$ is fixed in terms of the previously chosen normalization of the one-instanton term $C^{(1)}(x)$. (A possible constant term from integrating (\ref{eq:cnp2}) would add a multiple of $C^{(1)}(x)$ to  $C^{(2)}(x)$, which is excluded because it has a different exponential grading in the trans-series.)

The result (\ref{eq:cnp22}) for the two-instanton term $C^{(2)}(x)$ also exhibits resurgence properties. First, it is expressed explicitly in terms of the formal perturbative series $C^{(0)}(x)$. Second, we observe the generic  large-order/low-order resurgence relation between the large-order growth of the coefficients of the fluctuations about the one-instanton term, as shown in (\ref{eq:large-order-np1}), and the fluctuations of $C^{(2)}(x)$ about the $\xi(x)^2$ factor. These latter fluctuations can be generated from (\ref{eq:cnp22}) using the formal expansion of $C^{(0)}(x)$:
\begin{eqnarray}
C^{(2)}(x)\sim \xi(x)^2 \frac{e^{-2}}{2\pi} \left[\frac{1}{x}- 5 -\frac{11}{2} x -\frac{97}{2} x^2-\dots \right]
\label{eq:c2np-exp}
\end{eqnarray} 
Note the correspondence of the expansion coefficients in (\ref{eq:c2np-exp}) with the coefficients of the large-order growth of the one-instanton fluctuation term in (\ref{eq:large-order-np1}): $\left[1, -5, -\frac{11}{2}, -\frac{97}{2}, \dots\right]$.

\subsection{All Instanton Orders}

Remarkably, the  structure of expressions (\ref{eq:c1f}) and (\ref{eq:cnp22}), expressed in terms of the bivariate function $f(x, y)$, generalizes to {\it all orders} of the trans-series expansion (\ref{eq:c-trans}).
Expanding in powers of the trans-series ``instanton-counting'' parameter $\sigma$, we find that at order $k\geq 2$ we have a linear inhomogeneous  ODE for $C^{(k)}(x)$ of the form:
\begin{eqnarray}
\left(\frac{C^{(k)}}{C^{(1)}}\right)^\prime = F\left(C^{(0)}(x), C^{(1)}(x), C^{(2)}(x), \dots, C^{(k-1)}(x)\right)
\label{eq:cnpk}
\end{eqnarray}
This implies that $C^{(k)}$ can be expressed as a $(k-1)$-fold nested integral involving the lower order terms. This is again a manifestation of  resurgence: the $k$-instanton terms are expressed explicitly in terms of the lower instanton terms, all ultimately in terms of $C^{(0)}(x)$.

In fact, the situation is even more elegant than this. There are again identities in terms of the bivariate function $f(x, y)$ which permit all these nested integrals to be done, and we find a very simple all-orders trans-series expression:
\begin{align}
\begin{aligned}
C(x)&=C^{(0)}(x)+\sum_{k=1}^\infty \frac{\left(\sigma \cdot \xi(x)\right)^k}{k!}  \left[\left( f(x, y)\, \frac{\partial}{\partial y}\right)^{k-1} f(x, y)\right]_{y=C^{(0)}(x)}
\\
&= \left[ \exp \left(\sigma \xi(x) f(x, y)\, \frac{\partial}{\partial y} \right) \cdot y \right]_{y=C^{(0)}(x)}
\label{eq:final-trans}
\end{aligned}
\end{align}
This is the all-orders trans-series expansion of the solution to the non-linear ODE (\ref{eq:c}), giving the full non-perturbative solution of the Dyson-Schwinger equations of the massless Yukawa theory. Notice that each instanton order of the trans-series is expressed in terms of the formal perturbative series $C^{(0)}(x)$. We will prove this all-order result more explicitly in Section~\ref{sec:alien} using a completely different approach based on an explicit realization of \'Ecalle's alien calculus \cite{ecalle1981fonctions}.

\subsection{Trans-asymptotics: summing all instanton orders}
\label{sec:trans-asymptotics}

As $x\to +\infty$, the instanton factor $\xi(x)$ in (\ref{eq:inst}) is no longer small, and at a certain point the trans-series should be graded differently; as an expansion in powers of $x$, multiplied by functions of $\xi(x)$, rather than the other way around \cite{costin2008asymptotics,costin1999}.
To implement this trans-asymptotic matching, we re-arrange the trans-series (\ref{eq:c-trans}) as
\begin{eqnarray}
C(x)=x\sum_{n=0}^\infty x^n F_n\left(\rho(x)\right) 
\label{eq:xi-trans}
\end{eqnarray}
where the argument $\rho(x)$ is related to the instanton factor (\ref{eq:inst}) as
\begin{eqnarray}
\rho(x) \equiv \sigma \frac{\xi(x)}{x}
\end{eqnarray}
This means that for each order of the $x$ expansion we re-sum {\it all orders} of the instanton expansion in powers of $\sigma \frac{\xi(x)}{x}$. 
Inserting the expansion ansatz (\ref{eq:xi-trans}) into the Dyson-Schwinger equation (\ref{eq:c}) and substituting $\rho = \sigma \frac{\xi(x)}{x}$, the coefficient of $x$ gives the following equation for $F_0(\rho)$
\begin{eqnarray}
F_0(\rho) \rho\, \frac{d F_0}{d\rho}=-1+F_0(\rho)
\label{eq:F0}
\end{eqnarray}
This has the general solution
\begin{eqnarray}
F_0(\rho)=1+W(e^{c} \rho)
\label{eq:F0-sol}
\end{eqnarray}
in terms of the Lambert W function \cite{corless1996lambertw} which satisfies the defining equation $W(\rho) e^{W(\rho)}=\rho$, with $c$ an arbitrary parameter. The arbitrary parameter $c$ can be absorbed into the trans-series parameter $\sigma$ and the branch of $W$ is fixed by matching the expansion (\ref{eq:xi-trans})  to the trans-series structure in (\ref{eq:c-trans}) and (\ref{eq:final-trans}): 
\begin{align}
\begin{aligned}
C(x)&= C^{(0)}+\rho(x) x C^{(1)}(x) +\rho(x)^2 x^2 C^{(2)}(x)  +\rho(x)^3 x^3 C^{(3)}(x) +\dots \\
&= (C^{(0)}_0 x+\bigO(x^2))+\rho(x)(C^{(1)}_0 x+\bigO(x^2))+\rho(x)^2 (C^{(2)}_0 x+\bigO(x^2))\\%
&= x\left[C^{(0)}_0+C^{(1)}_0 \cdot \rho(x)+C^{(2)}_0 \cdot \rho(x)^2 +\dots \right] +x^2 \left[C_1^{(0)}+ C_1^{(1)}\cdot \rho(x)+\dots \right]+ \ldots  \\
&=x F_0(\rho(x))+x^2 F_1(\rho(x))+\dots 
\end{aligned}
\end{align}
Thus, the solution in (\ref{eq:F0-sol}) involves the $W_0$ branch, which is real and positive on the positive real line.
At the next order in $x$, we find the following first-order linear inhomogeneous ODE for $F_1$:
\begin{eqnarray}
F_0(\rho)\,\rho\frac{d}{d\rho} F_1(\rho)+\left(\rho \frac{d F_0}{d\rho}\right) F_1(\rho)-F_1(\rho)=3 F_0(\rho) \rho\frac{d F_0}{d\rho}-F_0^2(\rho)
\label{eq:F1}
\end{eqnarray}
This can be integrated in closed form, and matching to the re-arranged trans-series (\ref{eq:final-trans}) we obtain:
\begin{eqnarray}
F_1(\rho)=-\left(\frac{W^3(\rho)+3W(\rho)-1)}{2(1+W(\rho))}\right)
\label{eq:F1-sol}
\end{eqnarray}
This suggests that the $F_n(\rho)$ are rational functions of $W(\rho)$. This can be confirmed by changing variable  from $\rho$ to $W$, using the fact that 
\begin{eqnarray}
\rho\frac{d}{d\rho}=\frac{W}{1+W}\frac{d}{dW}
\label{eq:xi-W}
\end{eqnarray}
The ``doubly-resummed'' trans-asymptotic expansion therefore has the form
\begin{eqnarray}
C(x)=x\sum_{n=0}^\infty x^n {\mathcal F}_n\left(W\left(\frac{\sigma\, \xi(x)}{x} \right)\right)
\label{eq:W-trans}
\end{eqnarray}
in which all the terms ${\mathcal F}_n\left(W\right)$ are  rational functions of $W$. In other words, at each order $n$ of the perturbative expansion, all orders of the instanton expansion can be summed to give a rational function of the Lambert function $W\left(\frac{\sigma \xi(x)}{x} \right)$, which  is itself already an all-orders resummation of instanton terms. 

We comment that in certain integrable ODEs such as the Painlev\'e equations, the corresponding trans-asymptotic expansions are expressed in terms of rational functions, ${\mathcal G}_n(\sigma\, \xi(x)/x)$, of the one-instanton term $\xi(x)$ \cite{costin2008asymptotics,costin1999}. Interestingly, the Lambert W-function also appears in the solution of the massless Wess-Zumino model \cite{Bellon:2016mje}, in a 
non-commutative scalar QFT \cite{Panzer:2018tvy}, and in the context of algebraic group topology \cite{Borinsky:2019rtu}.  By its very definition,  the W-function is indeed natural in the transition between expansions in powers of couplings and in powers of instantonic exponentials \cite{Gardi:1998qr}. The Lambert function has two real branches; it would be interesting to study the analytic continuation to the second branch in the context of trans-asymptotics and the analyticity properties of the associated QFT Green's functions.

\section{Trans-series from the Alien Derivation for Formal Series}
\label{sec:alien}

\subsection{Alien Derivative Operator on the Ring of Formal Power Series}

The coefficients of the asymptotic expansion in eq.~\eqref{eq:large-order} have been evaluated in \cite{borinsky2018generating} using an approach that is entirely based on rings of power series. This approach to resurgence is a specialization of {\'E}calle's \cite{ecalle1981fonctions} general theory (see also \cite{Marino:2012zq,mitschi2016divergent,Aniceto:2018bis} where different aspects of this theory are highlighted), which does not require any information on the Borel transformation of the power series under inspection.

At the center of the approach from \cite{borinsky2018generating} is the \emph{alien} or \emph{asymptotic} derivative operator. 
Take the subspace $\mathbb{R}[[x]]^A_\beta \subset \mathbb{R}[[x]]$ of all formal power series $f(x) = \sum_{n=0} f_n x^n$ whose {\it coefficients} obey an asymptotic expansion of the form
\begin{align}
f_n \sim \sum_{k=0}^{\infty} c_k A^{n+\beta-k} \Gamma(n+\beta-k)
\text{ as } n\rightarrow \infty.
\label{eq:fringdef}
\end{align}
To each such power series $f(x)$ we can associate a new formal power series by interpreting the coefficients $c_k$ of the asymptotic expansion as a formal power series again. In this way we obtain a linear operator $\mathcal{A}^A_\beta: \mathbb{R}[[x]]^A_\beta \rightarrow \mathbb{R}[[x]]$, 
\begin{align}
\label{eqn:AsyDef}
( \mathcal{A}^A_\beta f )(x) := \sum_{k = 0}^\infty c_k x^k,
\end{align}
with the coefficients $c_k$ as in eq.~\eqref{eq:fringdef}. This operator is also often denoted as $\Delta_{A^{-1}}$, and called \emph{alien derivative}.

A simple but important example is the formal power series 
\begin{align}
I(x) = \sum_{n=0}^\infty (2n-1)!! x^n = \sum_{n=0}^\infty \frac{2^{n+\frac12} \Gamma(n+\frac12)}{\sqrt{2 \pi}} x^n 
\in \mathbb{R}[[x]]^2_{\frac12}
\end{align}
for which $( \mathcal{A}^2_{\frac12} I )(x) = \frac{1}{\sqrt{2 \pi}}$, because of the definitions in eqs.~\eqref{eq:fringdef} and \eqref{eqn:AsyDef}. This example is important for our discussion because $I(x)$ is the generating function for {\it all chord diagrams}. It is related to our function of interest $C(x)$, the generating function of {\it connected chord diagrams}, via the functional equation:
\begin{eqnarray}
I(x) = 1+ C(xI^2(x))
\label{eq:fun}
\end{eqnarray}
It is quite easy to prove this functional equation which $C$ fulfills \cite{Flajolet2000}. This functional equation can be used to solve the connection problem for the differential equation \eqref{eq:c} completely.

To do this, we need to know that the $\mathcal{A}^A_{\beta}$ operator is a \emph{derivation}, which respects the Leibniz and chain rules \cite{borinsky2018generating}:
\begin{align}
(\asy^A_{\beta} (f \cdot g)) (x) &= (\asy^A_{\beta} f) (x) g(x) + f(x) (\asy^A_{\beta} g) (x) 
\label{eq:leibniz} \\
(\asy^A_{\beta} (f \circ g)) (x) &= f'(g(x)) (\asy^A_{\beta} g)(x) + \left(\frac{x}{g(x)} \right)^{{\beta}}  e^{\frac{\frac{1}{x} - \frac{1}{g(x)}}{A}} (\asy^A_{\beta} f ) (g(x)), 
\label{eq:chainrule}
\end{align}
For the problem at hand, we can specialize these rules to $A = 2$ and $\beta = \frac12$. 

\subsection{Trans-series from the Functional Equation}

Since $\mathcal{A}^2_{\frac12}$ satisfies the Leibniz and chain rules, we can apply it directly to both sides of the functional equation (\ref{eq:fun}) to give an explicit expression for the $\mathcal{A}^2_{\frac12}$ derivative of $C$. We can therefore solve for the \emph{asymptotics} of the implicitly defined generating function $C$ in eq.~\eqref{eq:fun} by using $\mathcal{A}^2_{\frac12}$ in a similar way as we can solve for the ordinary derivative of an implicitly defined function (see \cite{borinsky2018generating} for details):
\begin{align}
    \label{eq:asymptotic_C}
    (\asy^{2}_{\frac12} C)(x) = \frac{1}{\sqrt{2\pi}} \frac{x}{C(x)} e^{-\frac{C(x)( C(x) + 2 )}{2x} },
\end{align}
Notice that the overall normalization of $(\asy^{2}_{\frac12} C)(x)$ is fixed by the functional equation (\ref{eq:fun}).
This result (\ref{eq:asymptotic_C}) should be compared with the one-instanton fluctuation term in eq.~\eqref{eq:c1}. With our chosen normalization of $C^{(1)}(x)$ in (\ref{eq:c1}), we have the exact identification:
\begin{eqnarray}
(\asy^{2}_{\frac12} C)(x) =  \sum_{n=0}^\infty C_n^{(1)} x^n
\label{eq:iden}
\end{eqnarray}
Therefore from eq.~\eqref{eq:fringdef} we have proved that the coefficients of $C(x) = \sum_{n=0}^\infty C_n^{(0)} x^n$ have the asymptotic expansion:
\begin{align}
    C_n^{(0)}  \sim \sum_{k=0}^\infty C^{(1)}_k 2^{n+\frac12-k} \Gamma\left(n+\frac12-k\right)\quad   \text{ as } n \rightarrow \infty.
\end{align}
Note that here we are directly working with the asymptotic behaviour of the original sequence and there is no arbitrary normalization constant anymore. The Stokes constant is fixed completely by the application of the chain rule (\ref{eq:chainrule}) and the functional equation (\ref{eq:fun}).

An important consequence of this observation is that it is now straightforward to iterate the $\asy^{2}_{\frac12}$ operator to calculate the asymptotic expansion of the asymptotic expansion, and so on. By using eq.~\eqref{eq:asymptotic_C} and the chain rule in eq.~\eqref{eq:chainrule}, we immediately get
\begin{align}
(\asy^{2}_{\frac12} (\asy^{2}_{\frac12} C) )(x) &= \frac{1}{\sqrt{2\pi}}
    \left( -\frac{1}{C(x)} - \frac{1}{x} - \frac{C(x)}{x} \right)
    \frac{x}{C( x)} e^{-\frac{C(x) ( C(x) + 2) }{2x}} (\asy^{2}_{\frac12} C)(x)
\nonumber
\\
&=
-\frac{1}{2\pi} \left( \frac{x^2}{C^3(x)} + \frac{x}{C^2(x)} + \frac{x}{C(x)}\right)
e^{-\frac{C(x) ( C(x) + 2) }{x}}
\\
&\equiv
\sum_{n\geq -1} C^{(2)}_n x^n
\end{align}
This expression is now actually a Laurent series, as it starts with a $\frac{1}{x}$-term, which matches with the two-instanton fluctuation factor in (\ref{eq:c2np-exp}) --- again thanks to our convenient choice of prefactor normalization. With this choice,
we obtain the explicit form of eq.~(\ref{eq:large-order-np1}) without resorting to any numerical approximation to fix the overall constant:
\begin{align}
    C_n^{(1)} &\sim \sum_{k=-1}^{\infty} C^{(2)}_k 2^{n+\frac12-k} \Gamma\left(n+\frac12-k\right) \quad  \text{ as } n\rightarrow \infty.
\end{align}

\subsection{All Orders Generating Function}

In principle, we could continue iterating the $\asy^{2}_{\frac12}$ operator operator indefinitely and calculate higher and higher order $\asy$ derivatives of $C$. This suggests that we might be able to find a three variable generating function $g(\eta,x,y)$ such that
\begin{align}
    g(\eta,x,C(x)) = \sum_{k\geq 0} \frac{\eta^k \left(  \left( \asy^{2}_{\frac12}\right)^{k} C \right)(x)}{k!},
\end{align}
for all higher asymptotic (alien) derivatives of $C(x)$.

We consider $g(\eta,x,y)$ to be a function of the variables $\eta,x,y$, even though we are eventually only interested in the specialization $y \rightarrow C(x)$. By the chain rule property (\ref{eq:chainrule}), $g(\eta,x,C(x))$ fulfills the equation
\begin{align}
    \label{eq:asyfeq}
    \asy^{2}_{\frac12} g(\eta,x, C(x)) &= \frac{\partial g(\eta,x, y)}{\partial y} \Big|_{y=C(x)} (\asy^{2}_{\frac12} C)(x) = \frac{\partial g(\eta,x, y)}{\partial y} f(x,y) \Big|_{y=C(x)},
\end{align}
where $f(x,y) = \frac{1}{\sqrt{2\pi}} \frac{x}{y} e^{-\frac{1}{2x} \left( y(y+2) \right) }$ is the bivariate generating function defined previously in eq.~\eqref{eq:fxc}.
On the other hand, because of the way $g$ is defined as an exponential generating function, 
\begin{align}
    \asy^{2}_{\frac12} g(\eta,x, C(x)) = \sum_{k\geq 0} \frac{\eta^k \left(  \left( \asy^{2}_{\frac12}\right)^{k+1} C \right)(x)}{k!} 
     =
     \frac{\partial g}{\partial \eta} (\eta,x, C(x)).
\end{align}
This equation which translates between the alien derivative and the `trans-series parameter' $\eta$ is also called \'Ecalle's bridge equation \cite{ecalle1981fonctions}. We stress that the $\mathcal A$ operator gives a simple and completely explicit realization of the bridge equation for this problem.

Using the bridge equation and eq.~\eqref{eq:asyfeq} shows that $g(\eta,x,y)$ is the solution of a partial differential equation
\begin{align}
     \frac{\partial g(\eta,x,y)}{\partial \eta} 
=
f(x,y)
\frac{\partial g(\eta,x,y)}{\partial y} \, ,
\end{align}
with the boundary condition $g(0,x,y) = y$. This PDE is solved by the formal expression
\begin{align}
\label{eqn:expgenfun}
    g(\eta,x,y) = 
 \exp \left(\eta\, f(x, y)\, \frac{\partial}{\partial y} \right) \cdot y.
\end{align}
The all-order trans-series solution in eq.~\eqref{eq:final-trans} of the differential equation \eqref{eq:c} is recovered after specifying $y = C(x)$ and $\eta = \sigma \xi(x)$.

Moreover, the PDE can also be solved explicitly with
\begin{align}
    g(\eta,x,y) = q^{-1}\left( x, \eta + q(x,y) \right),
\end{align}
where 
\begin{align}
    q(x,y) := \int_{0}^{y} \frac{dy'}{f(x,y')},
\end{align}
and $q(x,q^{-1}(x,y)) = y$. We therefore see that the all order asymptotics that $g(\eta,x,y)$ generates are encoded entirely in the function $f(x,y)$.

The form of this PDE is universal for all problems $h(x) = \sum_{n=0}^\infty h_n x^n$ with the property we call `functional resurgence'. This means that there exists a bivariate function $f(x,y)$, with a finite radius of convergence in $x$ and $y$, such that (compare with (\ref{eq:asymptotic_C})):
\begin{align}
(\asy^A_\beta h)(x) = f(x,h(x)) \quad.
\end{align}
This relation explicitly encodes the way the original power series $h(x)$ `resurges' inside its own asymptotic expansion, via the  function $f(x, y)$. Under the assumption that there are no additional leading singularities in the Borel plane of the solution (as is guaranteed here by the fact that the solution fulfills a first-order ODE), the function $f(x,y)$ contains all the non-perturbative information of $h$. Moreover, this information is easily accessible via the exponential generating function in eq.~\eqref{eqn:expgenfun}.

\section{Conclusions}

Our analysis shows that for 
quantum field theories for which the associated Hopf algebra structure reduces the Dyson-Schwinger equations to a set of coupled non-linear ODEs \cite{Broadhurst:1999ys,Broadhurst:2000dq,Kreimer:2006ua,Yeats:2008zy,vanBaalen:2008tc,vanBaalen:2009hu}, resurgent asymptotic analysis can be used to compute the full non-perturbative trans-series structure of physical quantities such as anomalous dimensions and beta functions. We have shown that the trans-series ansatz approach agrees perfectly with the alien calculus approach, based on an explicit and efficient alien derivative operator developed in \cite{borinsky2018generating,borinsky2018graphs}. 
The simplest illustrative example of this procedure has been discussed here, the four dimensional massless Yukawa theory, for which the relevant equation is a single first order non-linear ODE (\ref{eq:c}), and the resulting non-perturbative completion is a natural extension of the pioneering high-order perturbative results of Broadhurst and Kreimer \cite{Broadhurst:1999ys,Broadhurst:2000dq,Kreimer:2006ua}. The Hopf-algebraic analysis of this model is sufficiently sophisticated to reveal a rich non-perturbative structure, in the form of integer-repeated Borel singularities on the negative real axis, arising from the iterative structure of the Dyson-Schwinger equation.
 In other theories the ODE could be higher order, such as in the six dimensional scalar theory studied in \cite{Broadhurst:1999ys,Broadhurst:2000dq,Kreimer:2006ua}, for which the trans-series and associated combinatorics is similar but much richer \cite{inprep}, or the Dyson-Schwinger equations could become a system of coupled non-linear ODEs \cite{vanBaalen:2008tc,vanBaalen:2009hu,Yeats:2008zy,Bellon:2008zz,Bellon:2016mje,Bellon:2018lwy,Bersini:2019axn}. From general results  \cite{costin1998,costin2008asymptotics}, the resurgent ODE analysis extends to these more general cases. There has been significant recent progress \cite{Kompaniets:2017yct,McKane:2018ocs,Gracey:2018ame} in understanding the high-order perturbative behaviour of scalar quantum field theories, and we hope the resurgent trans-series approach might provide a useful new perspective on the associated non-perturbative physics.
Our considerations could be extended by using a more sophisticated combinatorial treatment of the respective Dyson-Schwinger equation. A Mellin transformation based approach as pioneered in \cite{Yeats:2008zy} looks especially promising for that (see also \cite[Chap. 9]{yeats2017combinatorial} for a recent review and \cite{Courtiel:2019dnq} for explicit calculations of asymptotics using this approach). Moreover, there is ongoing work to find a combinatorial interpretation for the variety of related sequences that appear after application of the $\asy$ operator \cite{inprep_mahmoud}. These considerations might also be helpful to understand the resurgence behaviour of these interesting combinatorial objects.

\section{Acknowledgements}
This material is based upon work supported by the U.S. Department of Energy, Office of Science, Office of High Energy Physics under Award Number DE-SC0010339 (GD) and by the NWO Vidi grant 680-47-551  ``Decoding Singularities of Feynman graphs'' (MB). Much of this work was done during visits by both authors to Humboldt University in 2018 and 2019, and at the Les Houches Summer School in 2018. We thank David Broadhurst, Dirk Kreimer and Karen Yeats for discussions and suggestions.

\end{document}